\documentclass[seceq,preprint]{ptptex}

\usepackage{graphicx}
\usepackage{side}
\usepackage{axodraw}
 
\newcommand{\slal}[2]{{{#1}\hspace{-8pt}{/}}_{#2}}
\newcommand{\kslash}{k\kern-1ex /}
\newcommand{\pslash}{p\kern-1ex /}
\newcommand{\qslash}{q\kern-1ex /}
\newcommand{\lslash}{l\kern-1ex /}
\newcommand{\sslash}{s\kern-1ex /}
\newcommand{\Dslash}{{\cal D}\kern-1.5ex /}

\newcommand{\beqa}{\begin{eqnarray}}
\newcommand{\eeqa}{\end{eqnarray}}

\newcommand{\be}{\begin{equation}}
\newcommand{\ee}{\end{equation}}
\newcommand{\la}{\langle}
\newcommand{\ra}{\rangle}
\newcommand{\ben}{\begin{eqnarray}}
\newcommand{\een}{\end{eqnarray}}
\newcommand{\nn}{\nonumber}
\def\lsim{\raise0.3ex\hbox{$<$\kern-0.75em\raise-1.1ex\hbox{$\sim$}}}
\def\gsim{\raise0.3ex\hbox{$>$\kern-0.75em\raise-1.1ex\hbox{$\sim$}}}
\def\simgt{\rlap{\lower 3.5 pt\hbox{$\mathchar \sim$}}\raise 1pt \hbox {$>$}}
\def\simlt{\rlap{\lower 3.5 pt\hbox{$\mathchar \sim$}}\raise 1pt \hbox {$<$}}
\newcommand{\cont}{{\rm cont}}
\newcommand{\latt}{{\rm latt}}

\newcommand{\msbar}{{\overline {\rm MS}}}

\newcommand{\csw}{{c_{\rm SW}}}

\newcommand{\ce}{{\it c}_{\it E}}
\newcommand{\cb}{{\it c}_{\it B}}
\newcommand{\lqcd}{{\Lambda}_{\rm QCD}}
\newcommand{\pos}{{p^*}}
\newcommand{\qos}{{q^*}}
\newcommand{\qoss}{{q^*_s}}
\newcommand{\qosd}{{q^*_d}}
\newcommand{\lo}{{(0)}}
\newcommand{\nlo}{{(1)}}
\newcommand{\mplo}{{m_p^{(0)}}}

\newcommand{\mpl}{{m_{p2}}}
\newcommand{\mph}{{m_{p1}}}
\newcommand{\mpllo}{{m_{p2}^{(0)}}}
\newcommand{\mphlo}{{m_{p1}^{(0)}}}

\notypesetlogo                       
\preprintnumber[3cm]{UTHEP-494}

\protect{\markboth{S.~Aoki, Y.~Kayaba, Y.~Kuramashi and N.~Yamada}
{On-shell improvement of the massive Wilson quark action}}

\title{On-shell improvement of the massive Wilson quark
action\thanks{presented by Y.~Kuramashi}
}

\author{
S.~\textsc{Aoki}$^{\rm a,c}$,
Y.~\textsc{Kayaba}$^{\rm a}$,
Y.~\textsc{Kuramashi}$^{\rm a,b}$ and
N.~\textsc{Yamada}$^{\rm c}$
}

\inst{
$^a$Institute of Physics, University of Tsukuba,\\
Tsukuba, Ibaraki 305-8571, Japan\\
$^b$Center for Computational Sciences, University of Tsukuba,\\
Tsukuba, Ibaraki 305-8577, Japan\\
$^c$RIKEN BNL Research Center, Brookhaven National Laboratory,\\
Upton, NY 11973, USA
}

\abst{
We review a relativistic approach to
the heavy quark physics in lattice QCD
by applying a relativistic $O(a)$ improvement to the massive Wilson
quark action on the lattice.
After explaining how power corrections of $m_Q a$ can be avoided and 
remaining uncertainties are reduced to be of order 
$(a\Lambda_{\rm QCD})^2$, we demonstrate a determination of
four improvement coefficients in the action up to
one-loop level in a mass dependent way. We also show a perturbative determination of mass
dependent renormalization factors and $O(a)$ improvement coefficients
for the vector and axial vector currents.
Some preliminary results of numerical simulations are also presented.
}

\begin{document}

\maketitle

\section{Introduction}

Lattice QCD should allow quantitative predictions for the
heavy quark physics from first principles. 
Up to now, however, most approaches have based on the
nonrelativistic effective theory, with which
the continuum limit can not be taken\cite{nrqcd}. 
In this report we review a new relativistic approach to
the heavy quark physics in lattice QCD
developed in a series of 
publications\cite{akt,csw_m0,param,bilinear,bilinear_dwf,latt01,latt02,latt03}. 
 
We consider a relativistic $O(a)$ improvement 
to deal with the heavy quarks on the lattice\cite{akt}. 
We discuss cutoff effects by extending the on-shell improvement
programme\cite{sym1,sym2,onshell,alpha} from massless to massive case. 
An important finding is that in our formulation
leading cutoff effects of order $(m_Qa)^n$ are absorbed
in the definition of renormalization factors for the quark mass 
and the wave function. 
After removing the next-leading cutoff effects of
$O((m_Qa)^n a\lqcd)$ with four parameters in the quark action 
properly adjusted in a $m_Qa$ dependent way, we are left with at most
$O((a\lqcd)^2)$ errors.
We show a determination of the four parameters in the quark
action up to one-loop level for various improved gauge actions\cite{param}.
They are determined free from the infrared divergences, 
once their tree level values are correctly tuned
in the $m_Q a$ dependent way.

We also make the $O(a)$ improvement of the vector and axial vector
currents at the one-loop level\cite{bilinear}.
We give a general discussion about  
what kind of improvement operators 
are required from the symmetries allowed on the lattice, in which
the Euclidean rotational symmetry is violated
because of $m_Q a$ corrections.
We consider both the heavy-heavy and heavy-light cases.

This report is organized as follows.
In Sec.~\ref{sec:action} we consider a relativistic $O(a)$ improvement
to handle the heavy quarks on the lattice
avoiding large $m_Q a$ corrections. The four parameters in the quark
action are determined up to one-loop level in Sec.~\ref{sec:param}.
In Sec.~\ref{sec:bilinear} 
we explicitly show the $O(a)$ improvement of the axial vector current
up to one-loop level.
With the use of the $O(a)$ improved quark action 
and axial vector current, 
we make some numerical studies in quenched QCD focusing on 
restoration of the space-time symmetry.  
Their results are presented in Sec.~\ref{sec:numerical}. 
Our conclusions are summarized in Sec.~\ref{sec:conclusion}.

\section{On-shell improvement of the massive Wilson quark action}
\label{sec:action}

We consider a relativistic $O(a)$ improvement to
control $m_Q a$ corrections for the heavy quarks on the
lattice. The basic idea is to apply the on-shell improvement
programme\cite{sym1,sym2,onshell,alpha}, which has been developed
in the small mass case, to the heavy quarks 
on th lattice. 
This method allows us to obtain the physical quantities 
in the continuum limit   
without requiring harsh condition $m_Q a \ll 1$ that is not 
achievable in near future.

Before going into details, we first remark on the on-shell improvement.
As explicitly stated in Ref.~\citen{alpha}, the on-shell improvement is
meant to improve the correlation functions in which local
composite fields are separated by non-zero physical distances.
All on-shell quantities (particle energies, 
scattering amplitudes, normalized matrix
elements of local composite fields between particle states, etc.) 
are extracted from these correlation functions\cite{alpha}. 
It should be stressed that the on-shell quantities 
are not restricted to the spectral ones.

Let us consider general cutoff effects for the heavy quarks on
the lattice, where the heavy quark mass $m_Q$ is allowed to be much
heavier than $\Lambda_{\rm QCD}$.
Under the condition that $m_Q\gg \lqcd$ and $m_Qa\sim O(1)$,
we assume that the leading cutoff
effects are
\be
f_0(m_Q a)>f_1(m_Q a)a\lqcd>f_2(m_Q a)(a\lqcd)^2>\cdots,
\ee
where $f_i(m_Q a)$ ($i\geq 0$) are smooth and continuous 
all over the range of 
$m_Q a$ and have Taylor expansions at $m_Q a=0$ with
sufficiently large convergence radii beyond $m_Q a=1$,
taking $f_0(0)=0$ and $f_i(0)\sim O(1)$ for $i\geq 1$.
The essential point in this assumption on cutoff effects
is that the cutoff effects of $O(a\lqcd)$ in the chiral limit
are still $O(a\lqcd)$ even if the quark mass is increased.
This means that our power counting is not based 
on the nonrelativistic effective theory. 
To control the scaling
violation effects we want to remove the cutoff effects up to
$f_1(m_Q a)a\lqcd$ by adding the counter terms
to the lattice quark action with the on-shell improvement.
If $m_Q a$ is small enough,
the remaining $f_2(m_Q a)(a\lqcd)^2$ contributions can be removed by
extrapolating the numerical data at several lattice spacings 
to the continuum limit. Otherwise, in case of
sufficiently small lattice spacing, 
the $O((a\Lambda_{\rm QCD})^2)$ errors can be neglected.

We first search for the relevant counter terms
required in the on-shell improvement. 
Listed below are the allowed operators under the
requirement of the gauge, axis interchange and other various
discrete symmetries on the lattice, where
the chiral symmetry is not imposed.
According to the work of Ref.~\citen{sw}, all the operators
with dimension up to six are given by
\ben
{\rm dim.3:}&& {\cal O}_3(x)={\bar q}(x)q(x), \\ 
{\rm dim.4:}&& {\cal O}_{4}(x)={\bar q}(x)\slal{D}{} q(x),\\
{\rm dim.5:}&& {\cal O}_{5a}(x)={\bar q}(x)D_\mu^2 q(x),\\
            && {\cal O}_{5b}(x)=i{\bar q}(x)
                                \sigma_{\mu\nu}F_{\mu\nu} q(x),\\
{\rm dim.6:}&& {\cal O}_{6a}(x)={\bar q}(x)\gamma_\mu D_\mu^3 q(x),\\
            && {\cal O}_{6b}(x)={\bar q}(x)D_\mu^2 \slal{D}{} 
                                q(x),\\
            && {\cal O}_{6c}(x)={\bar q}(x)\slal{D}{} 
                                D_\mu^2 q(x),\\
            && {\cal O}_{6d}(x)=i{\bar q}(x)\gamma_\mu[D_\nu,
                                F_{\mu\nu}] q(x),\\
            && {\cal O}_{6e}(x)={\bar q}(x){\slal{D}{}}^{\,3} q(x),\\
            && {\cal O}_{6f}(x)={\bar q}(x)\Gamma q(x)
                                {\bar q}(x)\Gamma q(x),
\een  
where $\Gamma=1,\gamma_5,\gamma_\mu,\gamma_\mu\gamma_5,\sigma_{\mu\nu}$.
These operators lead to a following generic form of the quark
action on the isotropic lattice:
\be
S_q^{\rm imp}=\sum_x\left[ 
c_3{\cal O}_3(x)+c_{4}{\cal O}_{4}(x) 
+\sum_{i=a,b} c_{5i}{\cal O}_{5i}(x) 
+\sum_{i=a,\dots,f} c_{6i}{\cal O}_{6i}(x)+\cdots  
\right],
\label{eq:qimp_iso}
\ee
where $c_3,\dots,c_{6f}$ are functions of the bare gauge
coupling $g$ and the power corrections of $m_Qa$.
At $m_Q=0$ the contributions of dimension 6 operators are of order
$(a\Lambda_{\rm QCD})^2$, which are negligible for the $O(a)$ improvement.
However, we are now interested in the case of $m_Q\gg \lqcd$ and
$m_Qa\sim O(1)$.

We first point out that, regardless of the magnitude of $m_Q a$, 
the $m_Qa$ corrections to the quark mass term
and the kinetic term can be absorbed 
in the renormalizations of the quark mass $Z_m$ and the
wave function $Z_q$.
For the sake of convenience we choose $c_3=m_0$ and $c_4=1$.

In the next step we reduce the number of basis
operators with the aid of the classical field equations.
It is easily found that ${\cal O}_{5a}$, ${\cal O}_{6b}$, 
${\cal O}_{6c}$ and  ${\cal O}_{6e}$ can be 
related to the quark mass term
or the kinetic term. In the on-shell improvement these
operators are redundant and can be eliminated 
from the action of eq.~(\ref{eq:qimp_iso}).
The operator ${\cal O}_{5a}$, however, 
is used to avoid the species doubling and the value of its
coefficient $c_{5a}$ is given by hand.  

The remaining operators are
${\cal O}_{5b}$, ${\cal O}_{6a}$, ${\cal O}_{6d}$ and 
${\cal O}_{6f}$. 
It is easy to see that ${\cal O}_{6d}$
and ${\cal O}_{6f}$ are $O((\Lambda_{\rm QCD}a)^2)$,
and therefore are irrelevant for the $O(a)$ improvement.
The operator ${\cal O}_{5b}$ is the so-called clover term,
for which
the nonperturbative method to determine the coefficient 
$c_{5b}$ in the massless limit is already
established.\cite{alpha} \  
However, the contributions of $(m_Qa)^n{\cal O}_{5b}$ 
($n\geq 1$) cannot be neglected in the 
present condition that allows $m_Q a\sim O(1)$. 
For $O(a\Lambda_{\rm QCD})$ improvement
the coefficient $c_{5b}$ has to be adjusted in the mass dependent way.

Our main concern is ${\cal O}_{6a}$.
Under the condition that $\partial_0 q(x)\sim m_Q q(x)$ and 
$\partial_i q(x)\sim p_i q(x)$ with $m_Q\gg |p_i|$ and $m_Qa\sim O(1)$, 
we have to treat the time and space
components of ${\cal O}_{6a}$ in a different way:
They follow different power counting.
Actually, the contribution of the time component ${\bar q}(x)\gamma_0
D_0^3 q(x)$, which is $O((\Lambda_{\rm QCD}a)^2)$ in the massless limit,
is not negligible any more in the present condition.
It can be related to other lower dimensional operators 
by the equation of motion:
\ben
a^2{\bar q}(x)\gamma_0 D_0^3 q(x)&=&-\frac{1}{a}(m_Q a)^3\bar q(x)q(x)\nn\\
&&-(m_Q a)^2\bar q(x)\gamma_i D_i q(x)
\label{eq:reduct_o6}\\
&&+a(m_Q a)\bar q(x) D_i^2 q(x)+O((\Lambda_{\rm QCD}a)^2).\nn
\een
This relation tells us that the contribution of 
${\bar q}(x)\gamma_0 D_0^3 q(x)$ is expressed 
by the lower dimensional operators multiplied by the power corrections
of $m_Qa$, so that the coefficients of the space derivative terms in
${\cal O}_4$ and ${\cal O}_{5a}$ become different from those
of the time derivative terms.
On the other hand, the contribution of the space component 
of ${\cal O}_{6a}$ is to be $O((\Lambda_{\rm QCD}a)^2)$, which is
negligible for the $O(a)$ improvement.
It is essential to note that 
the Lorentz non-covariant terms like ${\cal O}_{6a}$ yield
the difference of magnitude between 
the time and space components due to finite $m_Qa$ corrections.

The generalization of 
the above argument to any operators with higher
dimensions makes the discussion more transparent.
Let us consider  an arbitrary operator with
$4+k$ dimension,
$a^k {\cal O}_{4+k}$, where we write the lattice spacing $a$ explicitly.
The operator ${\cal O}_{4+k}$ contains $l$ pairs of $\bar q$ and $q$ and
$n$ covariant derivatives $D_\mu$ with $4+ k = 3 \times l + n$.
Using the classical field equation, 
some (but not all) of covariant derivatives
can be replaced by the quark mass $m_Q$. For $l\ge 2$ 
the largest possible power of the scaling violation is 
$ (m_Q a )^n (a\Lambda_{\rm QCD})^{3l-4}$. 
Therefore the operators which contain four or more quarks
are irrelevant for the $O(a\Lambda_{\rm QCD})$ improvement.
All the relevant contributions come from the quark bilinear operators. 
With the aid of the classical field equations, they can be
reduced to
\ben
&& (m_Q a)^n a^{-1} {\bar q}(x)q(x)  \\
&& (m_Q a)^{n-1} {\bar q}(x)\gamma_0 D_0 q(x),\;\; 
(m_Q a)^{n-1} \sum_i {\bar q}(x)\gamma_i D_i q(x)\\
&& (m_Q a)^{n-2}a {\bar q}(x) D_0^2 q(x),\;\; 
(m_Q a)^{n-2}a \sum_i {\bar q}(x) D_i^2 q(x)\\
&& (m_Q a)^{n-2} a i\sum_i {\bar q}(x)\sigma_{0i}F_{0i} q(x),\;\;
(m_Q a)^{n-2} a i\sum_{ij} {\bar q}(x)\sigma_{ij}F_{ij} q(x),
\een
for $n \geq 0$. The time and space components of 
${\cal O}_{4}$ and ${\cal O}_{5a,5b}$ should be treated separately
in case of finite $m_Qa$, where
the space-time asymmetry reflects the contributions of the higher
dimensional operators that break the rotational symmetry.
Now we know that the seven operators are
needed for the $O(a\lqcd)$ improvement.
Since three coefficients among these seven operators
can be absorbed in $Z_m$, $Z_q$ and the Wilson parameter $r_t$ 
for the time derivative of ${\cal O}_{5a}$ as already explained,
the remaining four coefficients have to be actually tuned.

In conclusion, at all order of $m_Q a$,
the generic quark action is
written as
\ben
S_q^{\rm imp}&=&\sum_x\left[ m_0{\bar q}(x)q(x)
+{\bar q}(x)\gamma_0 D_0q(x)
+\nu \sum_i {\bar q}(x)\gamma_i D_i q(x)\right.\nn\\
&&\left.-\frac{r_t a}{2} {\bar q}(x)D_0^2 q(x)
-\frac{r_s a}{2} \sum_i {\bar q}(x)D_i^2 q(x)\right.\nn\\
&&\left.-\frac{ig a}{2}\ce \sum_i {\bar q}(x)\sigma_{0i}F_{0i} q(x)
-\frac{ig a}{4}\cb \sum_{i,j} {\bar q}(x)\sigma_{ij}F_{ij} q(x)
\right],
\label{eq:action_q}
\een
where we are allowed to choose $r_t=1$ and the four parameters
$\nu$, $r_s$, $\ce$ and $\cb$ are to be adjusted.
In general these parameters have the form that
$X=\sum_n X^{(n)}(g^2) (m_Q a)^n$ 
with $X=\nu$, $r_s$, $\ce$ and $\cb$, and
$X^{(0)}$ should agree with the one in the massless $O(a)$ improved
theory: $\nu^\lo=1$, $r_s^\lo=r_t=1$, $\ce^\lo=\cb^\lo=
c_{\rm SW}$.\cite{alpha} \ 
Note that $\nu = 1 + O( (m_Qa)^2)$ and $r_s = r_t + O( m_Qa)$ since
the space-time asymmetry arises from Lorentz non-covariant terms such as
${\cal O}_{6a}$ via the on-shell reduction of eq.(\ref{eq:reduct_o6})
accompanied by power corrections of $m_Qa$.
In brief the differences between $\nu$ and 1, $r_t$ and $r_s$, 
$\ce$ and $\cb$ reflect the contributions of 
Lorentz non-covariant terms with higher dimensions.

We find that a similar quark action for the heavy quarks has
been proposed in Ref.~\citen{fnal}. The important
difference is that the parameter
$r_s$ is redundant in their formulation, 
while it should be tuned in ours.  
In this sense our action is equal to theirs 
with a special choice of parameters.
The reason of discrepancy in the number of
relevant operators between in our formulation and
that in Ref.~\citen{fnal} is explained in detail in Ref.~\citen{akt}.   
We explicitly show in the next section that the
parameter $r_s$ actually needs to be adjusted to
reproduce the correct on-shell quark-quark scattering amplitude.

\section{Determination of the improvement parameters in the quark action}
\label{sec:param}

\subsection{Tree level}

The four improvement parameters in the quark action of eq.(\ref{eq:action_q})
are determined such that $O(a)$ cutoff effects in on-shell
quantities (particle energies, scattering amplitudes, normalized matrix
elements of local composite fields between particle states, etc.) are removed.
We employ the on-shell quark-quark scattering
amplitude to determine the parameters $\nu$, $r_s$, $\ce$ and $\cb$ at
tree level, which
are adjusted to reproduce the continuum form of
the scattering amplitude 
removing the $m_Qa$ corrections\cite{akt},
\ben
T &=&-g^2(T^A)^2{\bar u}(p^\prime)\gamma_\mu u(p)D_{\mu\nu}(p-p^\prime)
{\bar u}(q^\prime)\gamma_\nu u(q)\nn\\
&&-g^2(T^A)^2{\bar u}(q^\prime)\gamma_\mu u(p)D_{\mu\nu}(p-q^\prime)
{\bar u}(p^\prime)\gamma_\nu u(q)\nn\\
&&+O((p_i a)^2,(q_i a)^2,(p^\prime_i a)^2,(q_i^\prime a)^2),
\een
where $p$, $q$ denote the incoming quark momenta and $p^\prime$,
$q^\prime$ for the outgoing quark ones.
$D_{\mu\nu}$ denotes the gluon propagator.
This improvement procedure follows the previous work\cite{c_sw} that
determined the $\csw=\cb=\ce$ parameter up to one-loop level in the
massless case.
At the tree level the quark-quark-gluon vertex is written as
\ben
\left({\bar u}(p^\prime)\Lambda_0^\lo(p,p^\prime) u(p)\right)_\latt
&=&Z_q^\lo\left({\bar u}(p^\prime)i\gamma_0 u(p)\right)_\cont
+O((p_i a)^2,(p_i^\prime a)^2),
\label{eq:vtx_nrm_tree_t}\\
\left({\bar u}(p^\prime)\Lambda_k^\lo(p,p^\prime) u(p)\right)_\latt
&=&Z_q^\lo\left({\bar u}(p^\prime)i\gamma_k u(p)\right)_\cont
+O((p_i a)^2,(p_i^\prime a)^2),
\label{eq:vtx_nrm_tree_s}
\een
for 
\ben
\Lambda_0^\lo(p,p^\prime)
&=&i\gamma_0\cos\left(\frac{p_0+p^\prime_0}{2}\right)\
+r_t \sin\left(\frac{p_0+p^\prime_0}{2}\right)\nn\\
&&+\frac{\ce^\lo}{2} \cos\left(\frac{p_0-p^\prime_0}{2}\right)
\sum_l \sigma_{0l}\sin (p_l-p^\prime_l),\\
\Lambda_k^\lo(p,p^\prime)
&=&i\nu^\lo \gamma_k\cos \left(\frac{p_k+p^\prime_k}{2}\right)\
+r_s^\lo \sin \left(\frac{p_k+p^\prime_k}{2}\right)\nn\\
&&+\frac{\ce^\lo}{2} \cos\left(\frac{p_k-p^\prime_k}{2}\right)
\sigma_{k0}\sin (p_0-p^\prime_0)\nn\\
&&+\frac{\cb^\lo}{2} \cos\left(\frac{p_k-p^\prime_k}{2}\right)
\sum_{l\ne k} \sigma_{kl}\sin(p_l-p^\prime_l),
\een
where the spinor on the lattice is given by
\be
u(p)=\left( \begin{array}{c}\phi\\
\frac{\nu{\vec p}\cdot{\vec \sigma}}{N(p)}\phi
\end{array}\right)
+O((p_i a)^2),
\ee
with $N(p)=(-i){\rm sin}(p_0)+m_0+r_t(1-{\rm cos}(p_0))
+r_s\sum_i(1-{\rm cos}(p_i))$.
The $O(a)$ improvement condition yields\cite{akt}
\ben
\nu^\lo&=&\frac{\sinh(\mplo)}{\mplo},
\label{eq:nu_me}\\
r_s^\lo&=&\frac{\cosh(\mplo)+r_t \sinh(\mplo)}{\mplo}
-\frac{\sinh(\mplo)}{\mplo^2},
\label{eq:rs_me}\\
\ce^\lo&=&r_t \nu^\lo,
\label{eq:ce}\\
\cb^\lo&=&r_s^\lo,
\label{eq:cb}
\een
where $\mplo$ is the tree-level pole mass explained below.
It is now shown that the four parameters are uniquely determined from the
on-shell quark-quark scattering amplitude. This is
an evidence we actually 
need four improvement parameters in the
quark action of eq.(\ref{eq:action_q}).

It is instructive to show that the $\nu$ and $r_s$ parameters 
are also determined from the
quark propagator which is obtained by inverting the Wilson-Dirac operator 
in eq.(\ref{eq:action_q}),
\ben
S_q^{-1}(p)&=&i \gamma_0 {\rm sin}(p_0)
+\nu i \sum_i \gamma_i {\rm sin}(p_i) +m_0 \nn\\
&&+r_t (1-{\rm cos}(p_0))+r_s\sum_i(1-{\rm cos}(p_i)),
\label{eq:qprop}
\een
At the tree level the parameters are adjusted such that the 
above quark propagator reproduces the correct relativistic form\cite{akt}:
\be
S_q(p)=
\frac{1}{Z_q^\lo}
\frac{ -i \gamma_0 p_0- i \sum_i \gamma_i p_i + m_p^\lo}
{p_0^2+\sum_i p_i^2 +{m_p^\lo}^2}
+{\rm (no\;\; pole\;\; terms)}+O((p_i a)^2)
\label{eq:qprop_rel}
\ee
around the pole. $Z_q^\lo$ and $m_p^\lo$ are extracted with $p_i=0$,
\ben
m_p^\lo&=&{\rm log}\left|\frac{m_0+r_t+\sqrt{m_0^2+2r_t m_0+1}}
{1+r_t}\right|,
\label{eq:polemass_0}\\
Z_q^\lo&=&{\rm cosh}(m_p^\lo)+r_t {\rm sinh}(m_p^\lo).
\een
Imposing finite spatial momenta 
the parameter $\nu$ is determined by demanding 
the correct relativistic spinor structure 
on $S^{-1}_q(p)$ of eq.(\ref{eq:qprop}). 
Comparing the coefficients of $\gamma_0$ 
and $\gamma_i$ in the numerator we obtain
\ben
\nu^\lo&=&\frac{{\rm sinh}(m_p^\lo)}{m_p^\lo}.
\label{eq:nu_qp}
\een
The parameter $r_s$ is determined 
such that the correct dispersion relation is reproduced:
\ben
E^2=m_p^2+\sum_i p_i^2+O(p_i^4).
\label{eq:disp}
\een
The result is
\ben
r_s^\lo&=&\frac{{\rm cosh}(m_p^\lo)+r_t {\rm sinh}(m_p^\lo)}{m_p^\lo}
-\frac{{\rm sinh}(m_p^\lo)}{{m_p^\lo}^2}\\
&=&\frac{1}{m_p^\lo}(Z_q^\lo-\nu^\lo).
\label{eq:rs_qp}
\een
It should be noted that the values of 
$\nu^\lo$ and $r_s^\lo$ are exactly the 
same as those determined from the on-shell quark-quark scattering
amplitude.
This is not an accident: The correct relativistic form of quark
propagator yields the correct relativistic form of Dirac spinor
required in the calculation of matrix elements.
Since it is simpler to treat the conditions on the quark propagator,
we employ them to determine the one-loop contributions to $\nu$ and
$r_s$ in the next subsection. 

\subsection{One-loop level}

The one-loop contributions to the quark self-energy come from
the rainbow and the tadpole diagrams, which are written as
\ben
g^2\Sigma(p,m_0)
&=&g^2
\left[i\gamma_0 \sin p_0 B_0(p,m_0)
+\nu i \sum_i\gamma_i \sin p_i B_i(p,m_0)+C(p,m_0)\right].\nn
\een
Incorporating this contribution, the inverse quark propagator
up to the one-loop level is written as
\ben
S^{-1}_q(p,m)&=&i\gamma_0\sin p_0[1-g^2 B_0(p,m)]
+\nu i \sum_i\gamma_i\sin p_i[1-g^2 B_i(p,m)]+m\nn\\
&&+2 r_t\sin^2 \left(\frac{p_0}{2}\right)
+2 r_s\sum_i \sin^2 \left(\frac{p_i}{2}\right) 
-g^2{\hat C}(p,m),
\label{eq:qp_inv_1loop}
\een
where we redefine the quark mass as
\ben
m&=&m_0-g^2 C(p=0,m=0),\\
{\hat C}(p,m)&=&C(p,m)-C(p=0,m=0).
\een
With this definition 
the inverse quark propagator satisfies the on-shell condition  
for the massless quark up to the one-loop level : 
$S^{-1}_q(p_0=0,p_i=0,m=0)=0$. 

\begin{figure}[t]
\centerline{
\includegraphics[width=80mm,angle=0]{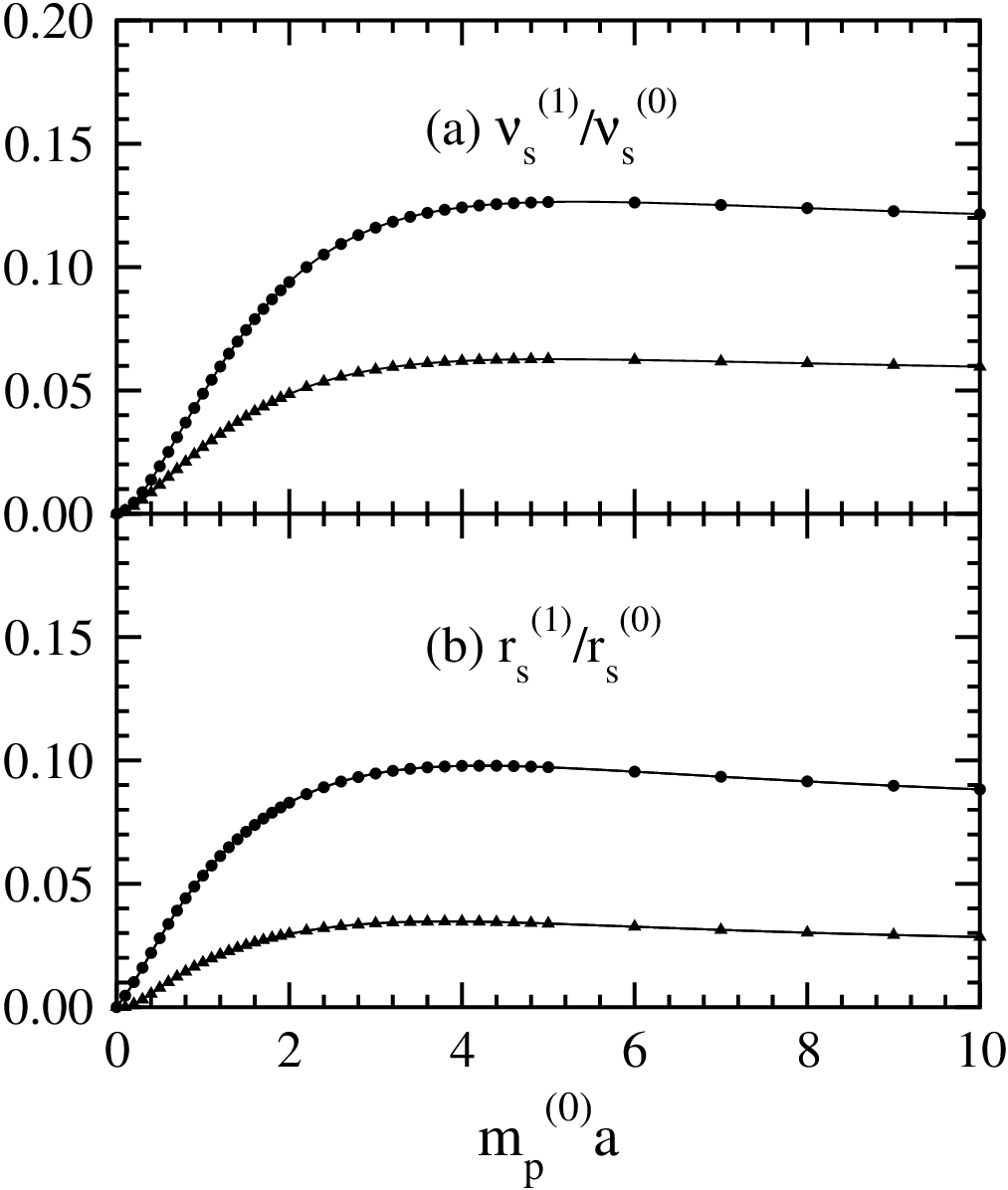}     
}
\caption{(a)$\nu^\nlo/\nu^\lo$ and (b)$r_s^\nlo/r_s^\lo$ 
as a function of $m_p^\lo$. Circles denote the plaquette gauge action
and triangles for the Iwasaki gauge action.}
\label{fig:nurs}
\vspace{8mm}
\end{figure}

The $\nu$ and $r_s$ parameters are determined by employing the same improvement
condition as the tree level.
The parameter $\nu$ is determined from
the relativistic spinor structure in $S^{-1}_q$ 
of eq.(\ref{eq:qp_inv_1loop}) at the pole. 
\ben
\nu[1-g^2 B_i(\pos,m)]=\frac{\sinh(m_p)}{m_p}[1-g^2 B_0(\pos,m)],
\een
where $\pos\equiv (p_0=im_p, p_i=0)$.
We show the quark mass dependences of $\nu^\nlo/\nu^\lo$ 
over the range $0\le \mplo\le 10$ for the plaquette and the 
Iwasaki gauge actions\cite{iwasaki} in Fig.~\ref{fig:nurs}(a).
The solid lines depict the results of the interpolation with a
rational expression. The errors are within symbols.
The parameter $r_s$ is determined from the relativistic dispersion
relation as done at the tree level.
The $\mplo$ dependences of $r_s^\nlo/r_s^\lo$ for the plaquette
and the Iwasaki gauge actions are shown 
in Fig.~\ref{fig:nurs}(b).

\begin{figure}[t]
\SetScale{0.6}
\begin{center}
\vspace{-10mm}

\begin{picture}(650,150)(0,0)
\ArrowLine(50,50)(75,75)
\Text(42,33)[l]{$p$}
\ArrowLine(75,75)(100,100)
\Vertex(75,75){2}
\ArrowLine(100,100)(75,125)
\Text(42,87)[l]{$q$}
\ArrowLine(75,125)(50,150)
\Vertex(75,125){2}
\Vertex(100,100){2}
\Gluon(100,100)(200,100){5}{8}
\Vertex(75,75){2}
\Vertex(75,125){2}
\Gluon(75,75)(75,125){5}{4}
\Text(75,18)[c]{(a)}
\ArrowLine(250,50)(275,75)
\Gluon(300,100)(275,75){5}{3}
\Vertex(275,75){2}
\Gluon(275,125)(300,100){5}{3}
\ArrowLine(275,125)(250,150)
\Vertex(275,125){2}
\Vertex(300,100){2}
\Gluon(300,100)(400,100){5}{8}
\Vertex(275,75){2}
\Vertex(275,125){2}
\ArrowLine(275,75)(275,125)
\Text(195,18)[c]{(b)}
\ArrowLine(450,50)(500,100)
\ArrowLine(500,100)(450,150)
\Vertex(500,100){2}
\GlueArc(520,100)(20,0,180){5}{7}
\GlueArc(520,100)(20,180,360){5}{7}
\Vertex(540,100){2}
\Gluon(540,100)(600,100){5}{5}
\Text(315,18)[c]{(c)}
\end{picture}

\vspace*{-15mm}

\begin{picture}(650,150)(0,0)
\ArrowLine(50,50)(100,100)
\ArrowLine(100,100)(50,150)
\Vertex(100,100){2}
\GlueArc(120,100)(20,-180,180){5}{14}
\GlueArc(150,100)(50,180,360){5}{16}
\Text(75,18)[c]{(d)}
\ArrowLine(250,50)(275,75)
\Vertex(275,75){2}
\ArrowArcn(275,100)(25,270,90)
\ArrowLine(275,125)(250,150)
\Vertex(275,125){2}
\GlueArc(275,100)(25,-90,90){5}{8}
\Gluon(400,75)(275,75){5}{11}
\Text(195,18)[c]{(e)}
\ArrowLine(450,50)(475,75)
\Vertex(475,75){2}
\ArrowArcn(475,100)(25,270,90)
\ArrowLine(475,125)(450,150)
\Vertex(475,125){2}
\GlueArc(475,100)(25,-90,90){5}{8}
\Gluon(475,125)(600,125){5}{11}
\Text(315,18)[c]{(f)}
\end{picture}

\vspace{-3mm}
\end{center}
\caption{Quark-quark-gluon vertices at one-loop level. 
$p$ ($q$) is incoming (outgoing) quark momentum.}
\label{fig:vtx_1loop}
\end{figure}
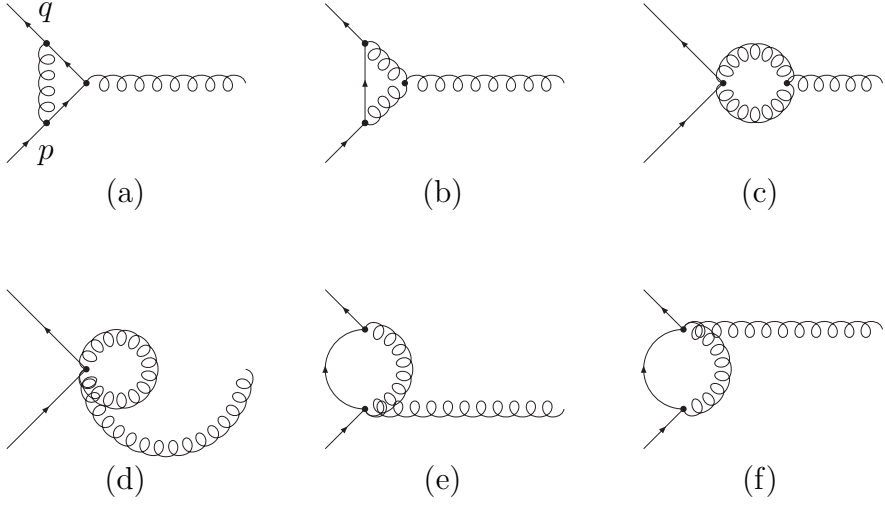

Let us turn to the one-loop calculation of $\ce$ and $\cb$. 
Recently the authors have shown the validity of 
the conventional perturbative method, which uses the fictitious gluon
mass as an infrared regulator\cite{gmass}, 
to determine the clover coefficient $\csw$ up to the one-loop level
in the massless case from the on-shell quark-quark
scattering amplitude\cite{csw_m0}.
We extend this calculation to the massive case.
According to Ref.~\citen{c_sw}, it is sufficient 
to improve each on-shell quark-quark-gluon vertex individually.
To determine the one-loop coefficients $\ce^\nlo$ and $\cb^\nlo$
we need six types of diagrams shown in Fig.~\ref{fig:vtx_1loop}.
We first consider to calculate $\cb^\nlo$.
Without the space-time symmetry the general form of the off-shell vertex
function at the one-loop level is written as
\ben
\Lambda_k^{(1)}(p,q,m) 
&=&\sum_{i=a,\dots,f}\Lambda_k^{(1-i)}(p,q,m)\nn\\
&=&\gamma_k F_1^k
+\gamma_k\{\pslash F_2^k+\pslash_0 F_3^k\}
+\{\qslash F_4^k+\qslash_0 F_5^k\}\gamma_k\nn\\
&&+\qslash\gamma_k\pslash F_6^k
+\qslash\gamma_k\pslash_0 F_7^k
+\qslash_0\gamma_k\pslash F_8^k
+\gamma_k\pslash_0\pslash F_9 
+\qslash\qslash_0\gamma_k F_{10}^k
\nn\\
&&+(p_k+q_k)\left[ H_1^k+\pslash H_2^k
+\qslash H_3^k+\qslash\pslash H_4^k\right]\nn\\
&&+(p_k-q_k)\left[ G_1^k+\pslash G_2^k
+\qslash G_3^k+\qslash\pslash G_4^k\right]+O(a^2),
\label{eq:vtx_k_offsh}
\een  
where $\Lambda_k(p,q,m)=\Lambda_k^\lo(p,q,m)
+g^2\Lambda_k^\nlo(p,q,m)+O(g^4)$ and  
$\pslash=\sum_{\alpha=0}^3 p_\alpha \gamma_\alpha$, 
$\qslash=\sum_{\alpha=0}^3 q_\alpha \gamma_\alpha$, 
$\pslash_0= p_0 \gamma_0$,
$\qslash_0= q_0 \gamma_0$.
The coefficients $F_i^k$ ($i=1,\dots,10$), $G_i^k$ ($i=1,2,3,4$) and 
$H_i^k$ ($i=1,2,3,4$) are functions of 
$p^2$, $q^2$, $p\cdot q$ and $m$.
 From the charge conjugation symmetry they have to satisfy
the following condition:
$F_2^k=F_4^k$,
$F_3^k=F_5^k$,
$F_7^k=F_8^k$,
$F_9^k=F_{10}^k$,
$H_2^k=H_3^k$,
$G_1^k=G_4^k=0$ and
$G_2^k=-G_3^k$.
Sandwiching $\Lambda_k^{(1)}(p,q,m)$ by the on-shell quark states
$u(p)$ and $\bar u(q)$, which satisfy
$\pslash u(p) = i m_p u(p)$ and $\bar u(q) \qslash = im_p \bar u(q)$,
the matrix element is reduced to 
\ben
&&{\bar u}(q)\Lambda_k^{(1)} (p,q,m)u(p)\nn\\
&=&{\bar u}(q)\gamma_k u(p)
\left\{F_1^k+i m_p (F_2^k+F_4^k)-m_p^2 F_6^k\right\} \nn\\
&&+(p_k+q_k){\bar u}(q)u(p)
\left\{H_1^k+im_p (H_2^k+H_3^k)-m_p^2 H_4^k\right\}\nn \\
&&+(p_k-q_k){\bar u}(q)u(p)
\left\{G_1^k+im_p (G_2^k+G_3^k)-m_p^2 G_4^k\right\}
+O(a^2),
\label{eq:vtx_k_onsh}
\een
where we use $F_3^k=F_5^k$,  $F_7^k=F_8^k$ and $F_9^k=F_{10}^k$.
(Note that we can replace $m_p$ with $\mplo$ in the one-loop diagrams.)
The first term in the right hand side contributes 
to the renormalization factor of the quark-quark-gluon vertex,
which is equal to $Z_q^\lo$ at the tree level.
With the use of $G_1^k=G_4^k=0$ and
$G_2^k=-G_3^k$, we find that the
last term of eq.(\ref{eq:vtx_k_onsh}) vanishes: 
this term is not allowed from the charge conjugation symmetry.
It is also possible 
to numerically check $G_1^k+i m_p (G_2^k+G_3^k)-m_p^2 G_4^k=0$.

The relevant term for the determination of $\cb$ is the third one,
which can be extracted by setting $p=\pos\equiv (p_0=im_p, p_i=0)$ and
$q=\qos\equiv (q_0=im_p, q_i=0)$ in eq.(\ref{eq:vtx_k_offsh}):
\ben
&&\left.H_1^k+im_p (H_2^k+H_3^k)-m_p^2 H_4^k\right|_{p=\pos,q=\qos}\nn\\
&=& \frac{1}{8}{\rm Tr} 
\left[\left\{\frac{\partial}{\partial p_k}
+\frac{\partial}{\partial q_k}\right\}\Lambda_k^\nlo(\pos,\qos,m)
(\gamma_0+1)\right]\nn\\
&&-\frac{1}{8}{\rm Tr}\left[\left\{\frac{\partial}{\partial p_i}
-\frac{\partial}{\partial q_i}\right\}
\Lambda_k^\nlo(\pos,\qos,m)(\gamma_0+1)\gamma_i\gamma_k\right]^{i\ne k},
\label{eq:order_a_s}
\een
where we have used the fact that $F^k$, $G^k$ and $H^k$ are functions of
$p^2$, $q^2$ and $p\cdot q$, so that
\ben
&&\left.\frac{\partial F_j^k}{\partial p_i}\right|_{p=\pos,q=\qos}
=\left.\frac{\partial F_j^k}{\partial q_i}\right|_{p=\pos,q=\qos}=0, \\
&&\left.\frac{\partial H_l^k}{\partial p_i}\right|_{p=\pos,q=\qos}
=\left.\frac{\partial H_l^k}{\partial q_i}\right|_{p=\pos,q=\qos}=0, \\
&&\left.\frac{\partial G_l^k}{\partial p_i}\right|_{p=\pos,q=\qos}
=\left.\frac{\partial G_l^k}{\partial q_i}\right|_{p=\pos,q=\qos}=0 
\een
with $j=1,\dots,10$, $l=1,2,3,4$ and $i=1,2,3$.

\begin{figure}[t]
\centering{
\hskip -0.0cm
\includegraphics[width=80mm,angle=0]{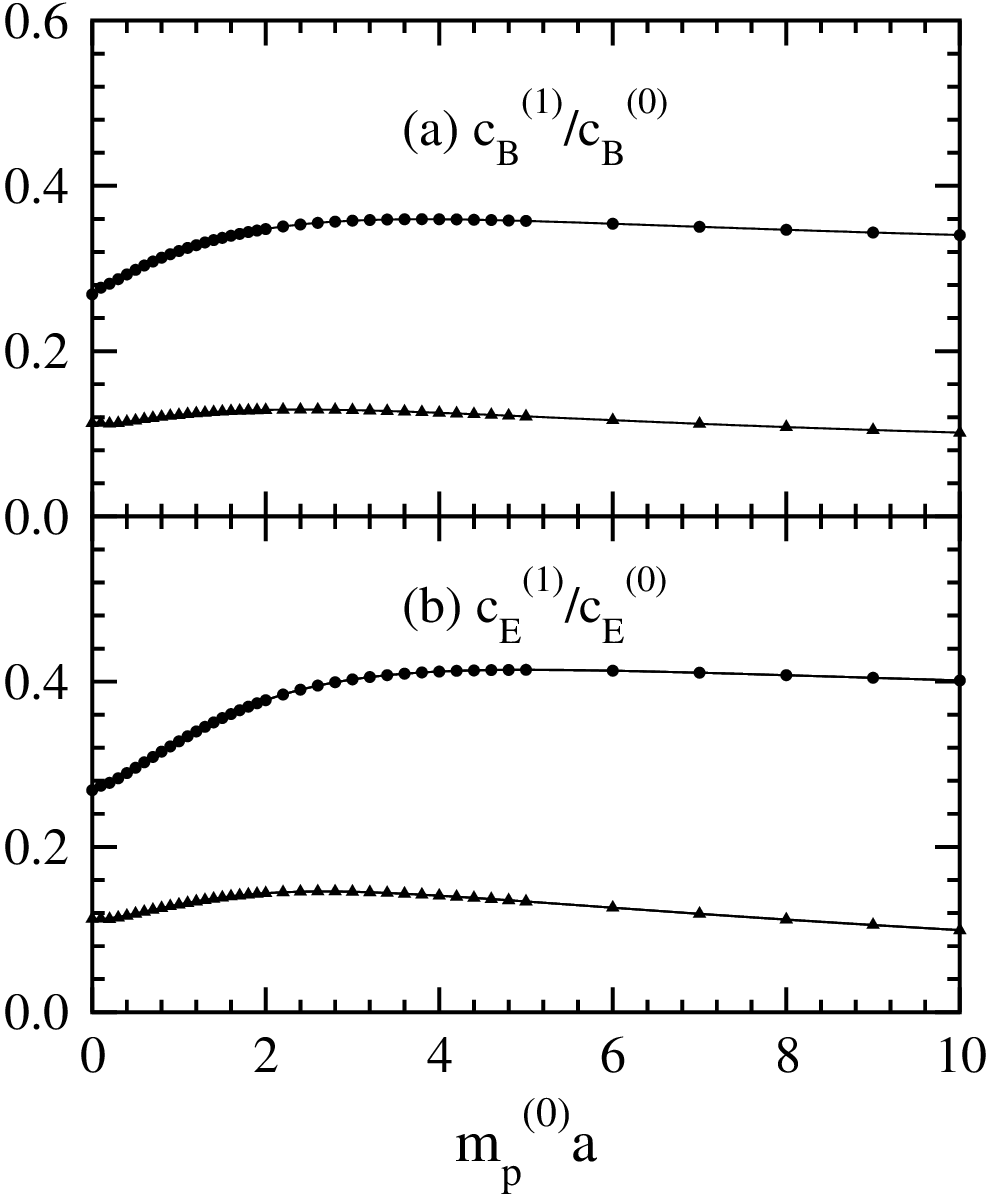}     
}
\caption{(a)$\cb^\nlo/\cb^\lo$ and (b)$\ce^\nlo/\ce^\lo$
as a function of $m_p^\lo$. Circles denote the plaquette gauge action
and triangles for the Iwasaki gauge action.}
\label{fig:cbce}
\vspace{8mm}
\end{figure}

We should remark that the third term in eq.(\ref{eq:vtx_k_onsh})
contains both the lattice artifact of $O(p_ka, q_ka)$ and
the physical contribution of $O(p_k/m,q_k/m)$.
The parameter $\cb$ is determined to eliminate 
the lattice artifacts of $O(p_ka, q_ka)$:
\ben
\frac{\cb^{(1)}-r_s^{(1)}}{2}&=&\left[H_1^k+i m_p (H_2^k+H_3^k)
-m_p^2 H_4^k\right]_{p=\pos,q=\qos}^\latt\nn\\
&&-Z_q^\lo\left[H_1^k+im_p (H_2^k+H_3^k)
-m_p^2 H_4^k\right]_{p=\pos,q=\qos}^\cont,
\een
where we take account of the tree-level expression for the
quark-quark-gluon vertex in
eq.(\ref{eq:vtx_nrm_tree_s}) and eq.(3.51) of Ref.~\citen{akt}.
We show the quark mass dependences of $\cb^\nlo/\cb^\lo$ 
over the range $0\le \mplo\le 10$ for the plaquette and the 
Iwasaki gauge actions in Fig.~\ref{fig:cbce}(a).
The solid lines represent the interpolation with 
a rational expression. The errors are within symbols.

The calculation for $\ce^\nlo$ is done in a similar way as for
$\cb^\nlo$: we determine $\ce^\nlo$ to remove the $O(a)$ contribution
from the on-shell matrix element ${\bar u}(q)\Lambda_0^{(1)}(p,q,m)u(p)$.
The quark mass dependences of $\ce^\nlo/\ce^\lo$ are also shown 
in Fig.~\ref{fig:cbce}(b).

Before closing this section
we should remark an important feature in the calculation 
of $\ce^\nlo$ and $\cb^\nlo$.
The infrared divergences originating 
from Figs.~\ref{fig:vtx_1loop}~(a), (b), (c) 
contain both the lattice artifacts and the 
physical contributions. The former exactly cancels out if and only if
the four parameters $\nu^\lo$, $r_s^\lo$, $\cb^\lo$ and $\ce^\lo$
are properly tuned. In Ref.~\citen{param} we demonstrate it by explicitly
writing down the infrared behaviors of the one-loop diagrams.  
This is another evidence that the tree-level improvement is correctly
implemented in Ref.~\citen{akt}.
It may be also instructive to mention the massless case\cite{c_sw,csw_m0}, 
where we find a similar situation:
The infrared divergences originating 
from Figs.~\ref{fig:vtx_1loop}~(a), (b), (c), (e), (f) 
cancel out if and only if $\csw^\lo$ is correctly adjusted
at the tree level, namely $\csw^\lo=1$.

\section{$O(a)$ improvement of the axial vector currents}
\label{sec:bilinear}

We consider the on-shell $O(a)$ improvement of the axial
vector current both for the heavy-heavy and heavy-light cases.
The discussion for the vector case is in parallel with the axial vector case.
(See Ref.~\citen{bilinear} for details.)

The renormalized operator with the $O(a)$ improvement is 
given by
\ben
A^{\latt,R}_\mu(x)&=&
Z_{A_\mu}^{\latt} \left[ 
{\bar q(x)} \gamma_\mu\gamma_5 Q(x)
-g^2 c_{A_\mu}^+
\partial_\mu^+ \{{\bar q(x)} \gamma_5 Q(x)\}
-g^2 c_{A_\mu}^- \partial_\mu^- \{{\bar q(x)}
\gamma_5 Q(x)\}\right.\nn\\
&&\left.+g^2 c_{A_\mu}^L \{{\vec
\partial_i}{\bar q(x)}\} \gamma_i 
\gamma_\mu\gamma_5 Q(x) 
-g^2 c_{A_\mu}^H {\bar q(x)}
\gamma_\mu\gamma_5 \gamma_i \{{\vec \partial_i} Q(x)\}+O(g^4)\right],
\label{eq:a_r}\
\een
where we assume that the Euclidean space-time rotational symmetry 
is not retained on the lattice. 
The coefficients $Z_{A_\mu}^{\latt}$ and $c_{A_\mu}^{(+,-,H,L)}$ 
are functions of
the quark masses $m_Q$ and $m_q$. 
With the aid of equation of motion
we are allowed to set $c^H_{A_0}=c^L_{A_0}=0$.
In the special case of $m_Q=m_q$, $c_{A_\mu}^-=0$ and
$c_{A_\mu}^H=-c_{A_\mu}^L$ are derived from the charge conjugation symmetry.
We also note that all the improvement coefficients
except $c_{A_\mu}^+$ vanish in the limit of $m_Q=m_q=0$.
We determine $Z_{A_\mu}^{\latt}$ 
and $c_{A_\mu}^{(+,-,H,L)}$ at the one-loop level for
both heavy-heavy and heavy-light cases.
In the following we are restricted 
to the time component of the axial vector current
because of the limitation of space. 
As for the space component refer to Ref.~\citen{bilinear}.

The general form of the off-shell vertex function
at the one-loop level on the lattice is given by
\ben
\Lambda_{05}^{(1)}(p,q,\mph,\mpl) 
&=&\gamma_0\gamma_5 F_1^{05}
+\gamma_0\gamma_5\pslash F_2^{05}
+\qslash \gamma_0\gamma_5 F_3^{05}
+\qslash\gamma_0\gamma_5\pslash F_4^{05}\nn\\
&&+(p_0-q_0)\left[ \gamma_5G_1^{05}+\gamma_5\pslash G_2^{05}
+\qslash\gamma_5 G_3^{05}+\qslash\gamma_5\pslash G_4^{05}\right]
\label{eq:a_0_s_offsh}\\
&&+(p_0+q_0)\left[ \gamma_5H_1^{05}+\gamma_5\pslash H_2^{05}
+\qslash\gamma_5 H_3^{05}+\qslash\gamma_5\pslash H_4^{05}\right]
+O(a^2),\nn
\een  
where $p$ denotes incoming quark momentum and $q$ for outgoing one. 
The coefficients $F^{0 5}$, $G^{0 5}$, $H^{0 5}$ are functions of 
$p^2$, $q^2$, $p\cdot q$, $\mph$ and $\mpl$.

Sandwiching eq.(\ref{eq:a_0_s_offsh}) by the on-shell quark states
$u(p)$ and $\bar u(q)$, which satisfy
$\pslash u(p) = i \mph u(p)$ and $\bar u(q) \qslash = i\mpl \bar u(q)$,
the matrix element is reduced to 
\ben
&&{\bar u}(q)\Lambda_{05}^{(1)}(p,q,\mph,\mpl)u(p)\nn\\
&=&{\bar u}(q)\gamma_0\gamma_5 u(p)
\left\{F_1^{05}+i\mph F_2^{05}+i\mpl F_3^{05}-\mph\mpl F_4^{05}\right\} \nn\\
&&+(p_0-q_0){\bar u}(q)u(p)
\left\{G_1^{05}+i\mph G_2^{05}+i\mpl G_3^{05}-\mph\mpl G_4^{05}\right\}
\label{eq:a_0_s_onsh}\\
&&+(p_0+q_0){\bar u}(q)\gamma_5u(p)
\left\{H_1^{05}+i\mph H_2^{05}+i\mpl H_3^{05}-\mph\mpl H_4^{05}\right\}
+O(a^2),\nn
\een
where the coefficients are summarized as
\ben
X_{05}&=&F_1^{05}+i\mph F_2^{05}+i\mpl F_3^{05}-\mph\mpl F_4^{05}, \\
Y_{05}&=&G_1^{05}+i\mph G_2^{05}+i\mpl G_3^{05}-\mph\mpl G_4^{05}, \\
Z_{05}&=&H_1^{05}+i\mph H_2^{05}+i\mpl H_3^{05}-\mph\mpl H_4^{05}. 
\label{eq:c_05_s_onsh}
\een

Since the above coefficients contain 
both the lattice artifacts and the physical contributions 
which remain in the continuum,
we have to isolate the lattice artifacts in order to determine 
the improvement coefficients in eq.(\ref{eq:a_r}).
The improvement coefficients are given by
\ben
\Delta_{\gamma_0\gamma_5}&=&\left(X_{05}\right)^\latt-\left(X_{05}\right)^\cont,
\label{eq:d_a_0}\\
ic_{A_0}^+&=&\left(Y_{05}\right)^\latt-\left(Y_{05}\right)^\cont,
\label{eq:+_a_0}\\
ic_{A_0}^-&=&\left(Z_{05}\right)^\latt-\left(Z_{05}\right)^\cont,
\label{eq:-_a_0}
\een
where the continuum contributions are obtained by 
employing the naive dimensional
regularization (NDR) with the modified minimal subtraction 
scheme ($\msbar$). 
Note that $c_{A_0}^-=0$ 
for $\mph=\mpl$ from the charge conjugation symmetry.

Here we should remark an important point. 
$(X_{05})^\latt,(Y_{05})^\latt,(Z_{05})^\latt$ are process-dependent:
They are allowed to have different finite values and infrared divergences,
which are regularized by the fictitious gluon mass in our calculation,
for decay and scattering processes.
However, once we subtract the continuum counter part,
the process dependences are canceled out and 
we are left with finite constants, namely the improvement
coefficients $\Delta_{\gamma_0\gamma_5},c_{A_0}^+,c_{A_0}^-$.

Combining $\Delta_{\gamma_\mu\gamma_5}$ and the wave function 
renormalization factors, we obtain
the renormalization factor of the axial vector currents:
\ben
\frac{Z_{A_\mu}^\latt}{Z_{A_\mu}^\cont}
=\sqrt{Z_{Q,\latt}^\lo(\mphlo)}
\sqrt{Z_{q,\latt}^\lo(\mpllo)}\left(1-g^2\Delta_{A_\mu}\right)
\label{eq:z_a}
\een
with
\ben
Z_{Q,\latt}^\lo(\mphlo)&=&\cosh(\mphlo)+r_t \sinh(\mphlo)\\
Z_{q,\latt}^\lo(\mpllo)&=&\cosh(\mpllo)+r_t \sinh(\mpllo)\\
\Delta_{A_\mu}&=&\Delta_{\gamma_\mu\gamma_5}
-\frac{\Delta_{Q}}{2}-\frac{\Delta_{q}}{2},
\een
where $\Delta_{Q,q}$ are found in Ref.~\citen{param}.

Employing a set of special momentum assignments $p=\pos\equiv (p_0=im_{p1}, p_i=0)$ and
$q=\qoss\equiv (q_0=im_{p2}, q_i=0)$ or 
$q=\qosd\equiv (q_0=-im_{p2}, q_i=0)$, where subscripts $s$ and $d$ represent
the scattering and the decay respectively,
we extract
the relevant coefficients $X_{05},Y_{05},Z_{05}$ for $A_0$ 
from the off-shell vertex function (\ref{eq:a_0_s_offsh}):
\ben
&&X_{05}^s
\label{eq:a_0_x}\\
&&=\frac{1}{4}{\rm Tr} \left[\Lambda_{05}^{(1)}\gamma_5\gamma_0
+i\mph \frac{\partial}{\partial p_k}
\Lambda_{05}^{(1)}(1+\gamma_0)\gamma_k\gamma_5
+i\mpl \frac{\partial}{\partial q_k}
\Lambda_{05}^{(1)}(1+\gamma_0)\gamma_k\gamma_5\right]_{p=\pos,q=\qoss},
\nn\\
&&X_{05}^s+i\mph(Y_{05}^s+Z_{05}^s)-i\mpl(Y_{05}^s-Z_{05}^s)\nn\\
&&=\frac{1}{4}{\rm Tr} \left[\Lambda_{05}^{(1)}\gamma_5(1+\gamma_0)
+2i\mph \frac{\partial}{\partial p_k}
\Lambda_{05}^{(1)}(1+\gamma_0)\gamma_k\gamma_5\right]_{p=\pos,q=\qoss},
\label{eq:a_0_xyz_s}\\
&&X_{05}^d-i\mph(Y_{05}^d+Z_{05}^d)-i\mpl(Y_{05}^d-Z_{05}^d)=
-\frac{1}{4}{\rm Tr} \left[\Lambda_{05}^{(1)}(1+\gamma_0)\gamma_5\right]_{p=\pos,q=\qosd},
\label{eq:a_0_xyz_d}
\een
where superscripts $s$ and $d$ in $X_{05},Y_{05},Z_{05}$ represent
their momentum assignments.
and $F^{05}$, $G^{05}$ and $H^{05}$ are functions of
$p^2$, $q^2$ and $p\cdot q$ resulting in
\ben
&&\left.\frac{\partial F^{0 5}}{\partial p_i}\right|_{p=\pos,q=\qos}
=\left.\frac{\partial F^{0 5}}{\partial q_i}\right|_{p=\pos,q=\qos}=0, \\
&&\left.\frac{\partial H^{0 5}}{\partial p_i}\right|_{p=\pos,q=\qos}
=\left.\frac{\partial H^{0 5}}{\partial q_i}\right|_{p=\pos,q=\qos}=0, \\
&&\left.\frac{\partial G^{0 5}}{\partial p_i}\right|_{p=\pos,q=\qos}
=\left.\frac{\partial G^{0 5}}{\partial q_i}\right|_{p=\pos,q=\qos}=0 
\een
with $i=1,2,3$.

\begin{figure}[t]
\centering{
\hskip -0.0cm
\includegraphics[width=80mm,angle=0]{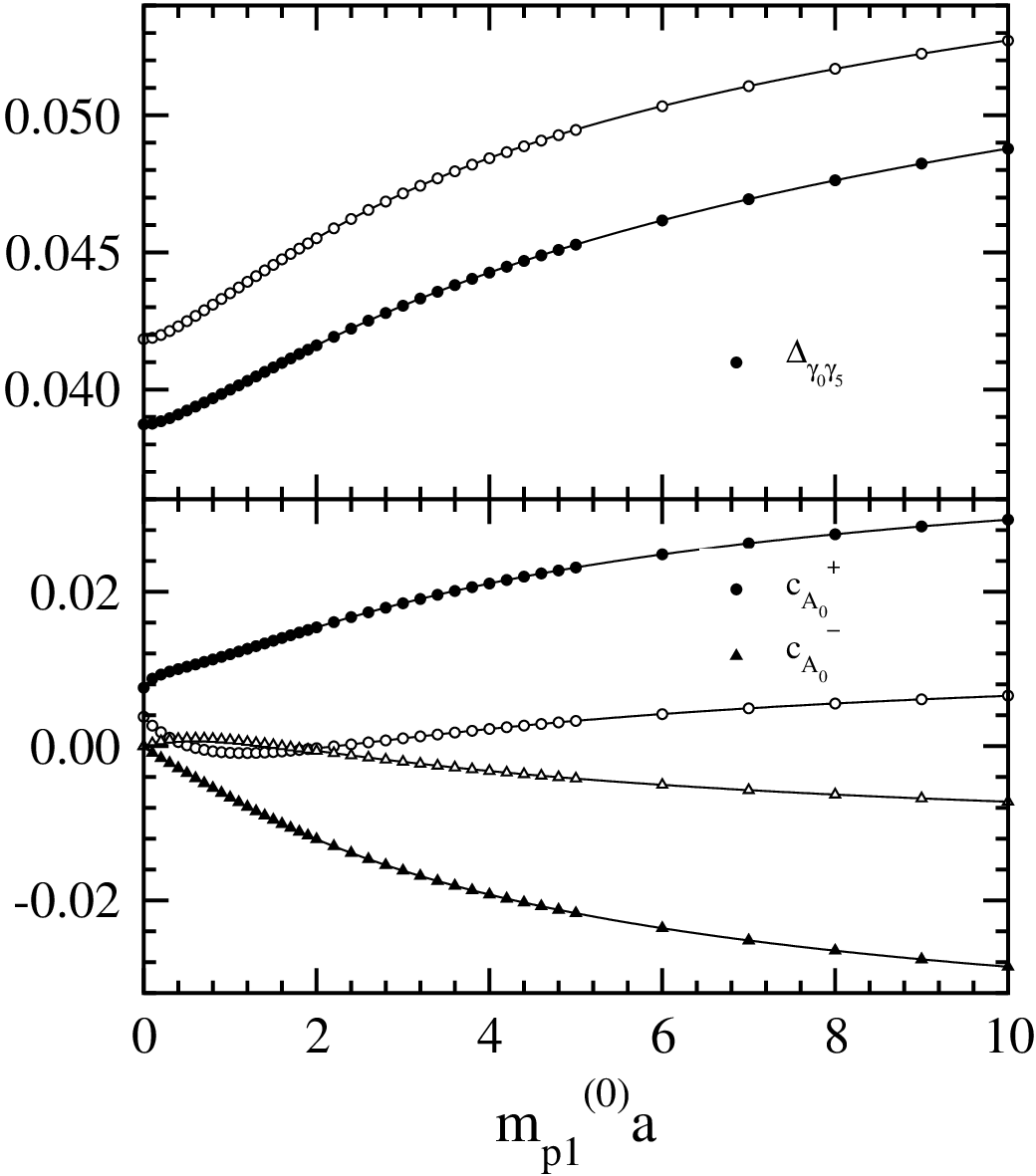}     
}
\caption{$\Delta_{\gamma_0\gamma_5}, c_{A_0}^{(+,-)}$ for heavy-light case as 
a function of $\mphlo$. Solid symbols denote the plaquette gauge action
and open ones for the Iwasaki gauge action.}
\label{fig:a_0_hl}
\vspace{8mm}
\end{figure}                                           


The set of improvement coefficients are determined from
eqs.(\ref{eq:d_a_0}$-$\ref{eq:-_a_0}).
Assuming that the $\mpl a$ corrections are negligible, 
we evaluate the improvement coefficients, 
except $c_{A_0}^{(+,-)}$, as a function of
$\mphlo$ with $\mpllo=0$.
In eqs.(\ref{eq:a_0_xyz_s}), (\ref{eq:a_0_xyz_d}) 
we find that $Y_{05}-Z_{05}$ are not determined
if we set $\mpllo=0$. 
Therefore one should extrapolate data at non-zero 
$\mpllo$ to $\mpllo=0$. We however keep $\mpllo=0.0001$ in our calculation to 
determine $c_{A_0}^{(+,-)}$ since the difference between
the value at $\mpllo=0.0001$ and the one extrapolated to $\mpllo=0$ is
less than 1 \%.
In Fig.~\ref{fig:a_0_hl} we show
numerical results of $\Delta_{\gamma_0\gamma_5}$, $c_{A_0}^{(+,-)}$ for
the heavy-light case.
The solid lines denote the interpolation with 
a rational expression. The errors are within symbols.
We can find the heavy-heavy case in Fig.~9 of Ref.~\citen{bilinear}. 

We also make a brief comment on our recent work of the $O(a)$
improvement for the heavy-light vector and axial vector currents with
relativistic heavy and domain-wall light quarks\cite{bilinear_dwf}.
The most important feature in this calculation is that
the renormalization and the improvement coefficients of the heavy-light
vector current agree with those of the axial vector current. We have
shown that this is indebted to the exact chiral symmetry for the light
quark irrespective of the heavy quark action.
We have also presented how to implement the on-shell improvement on the
massive domain-wall quark action, which is required to cancel out the
infrared divergences generating from the one-loop vertex corrections.

\section{Numerical studies in quenched QCD}
\label{sec:numerical}

Now we have achieved the $O(a)$ improvement both for the quark action
and the axial vector current, the next step is to check the
effectiveness of the improvement.
Our numerical studies are carried out in quenched QCD
focusing on the restoration of the space-time symmetry for the
heavy-heavy and heavy-light meson systems.  
We investigate the dispersion relation of moving mesons and
the difference of the pseudoscalar meson decay constants
extracted from the temporal and spatial components 
of axial vector currents. 

We take the clover action with non-perturbative $c_{\rm SW}^{\rm NP}$
\cite{csw_alpha,csw_cppacs} for light quarks.
As for the heavy quarks, we replace the massless contribution in 
$\ce^\nlo$ and $\cb^\nlo$ of the improved heavy quark action with
that of the non-perturbative one as
\ben
\cb= \{\cb^{\rm PT}(m_Q a) -\cb^{\rm PT}(0)\}
+c_{\rm SW}^{\rm NP},\\ 
\ce= \{\ce^{\rm PT}(m_Q a) -\ce^{\rm PT}(0)\}
+c_{\rm SW}^{\rm NP}, 
\een
where the superscript PT represents ''perturbative'' value. 
With this replacement $O(a)$ errors are completely removed at $m_Q=0$.
This is required by a consistency between the light and heavy quarks.
We generated 200 configurations with the plaquette 
and the Iwasaki gauge actions 
on a $24^3 \times 48$ lattice at $a^{-1} \approx 2$ GeV. 
We employ three light quark masses corresponding to
$m_{\rm PS}/m_{\rm V} \sim 0.56 - 0.77$ and four
heavy quark masses covering the charm quark mass.
For comparison we make another simulation 
with the clover action both for the heavy and light quarks.

\begin{figure}[t]
\centering{
\hskip -0.0cm
\includegraphics[width=100mm,angle=0]{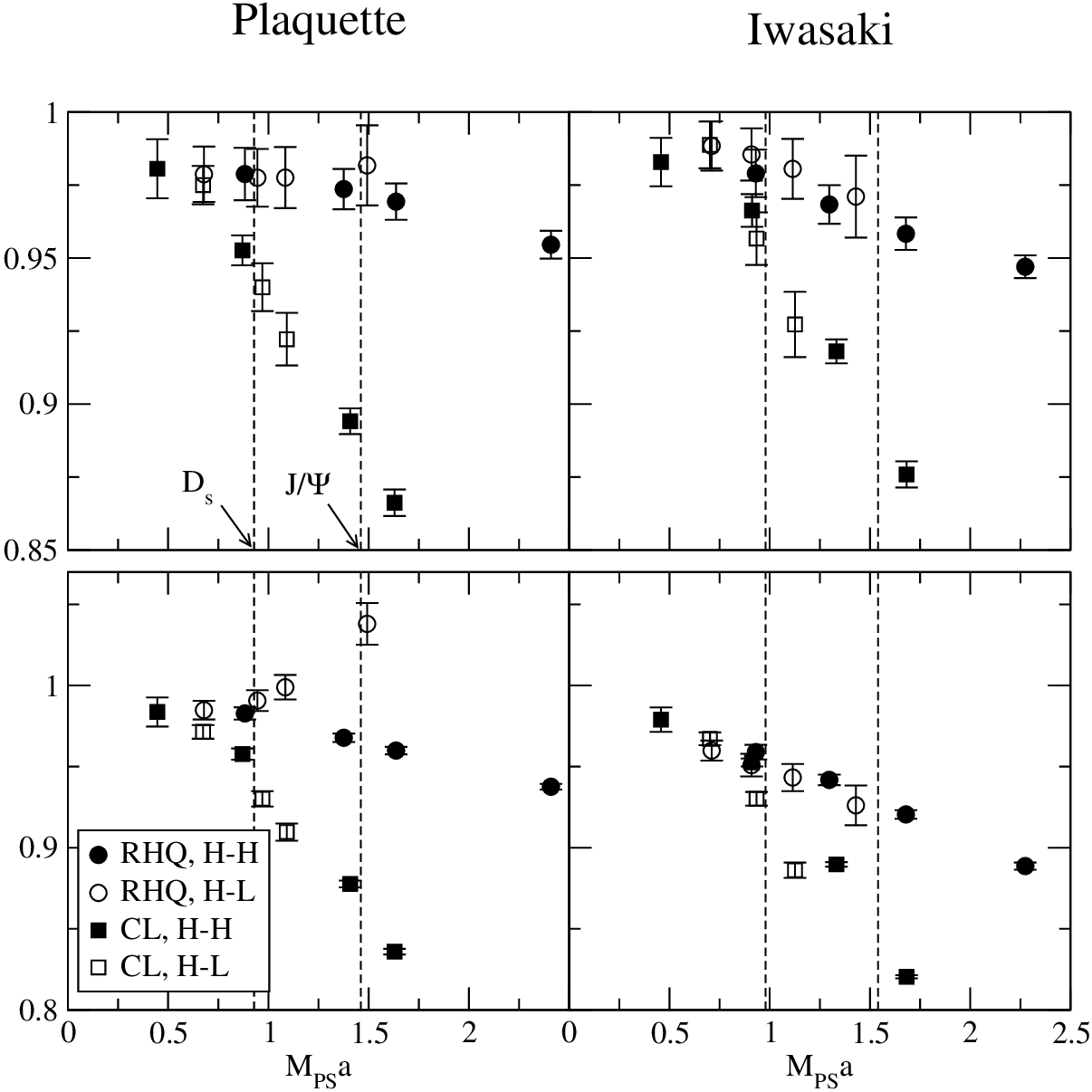}     
}
\caption{Effective speed of light(top) and pseudoscalar meson decay
constants as a function of the meson masses for the plaquette
and the Iwasaki gauge actions. RHQ denotes 
the $O(a)$ improved heavy quark action and CL represents 
the clover action for heavy quarks.}
\label{fig:splais}
\vspace{8mm}
\end{figure}                                           

To investigate the dispersion relation of 
heavy-heavy and heavy-light mesons,
it is convenient to define the effective speed of light as
\ben
c_{\rm eff}=\sqrt{\frac{E(p_s)^2-E(0)^2}{p_s^2}}, 
\een
where $E(p_s)$ is the pseudoscalar meson energy 
with the spatial momentum $p_s$.
This quantity is supposed to be unity in the continuum limit,
which means the restoration of relativistic dispersion relation.
In Figs.~\ref{fig:splais}(a), (b)
we plot numerical results
as a function of the meson mass for
heavy-heavy(H-H) and heavy-light(H-L) systems. 
We observe that around $J/\Psi$ mass in the heavy-heavy system 
the deviation from $c_{\rm eff}=1$
is equal to or larger than about 10\% for the clover heavy quark action, 
while less than 4\% for the $O(a)$ improved heavy quark action.

We also measure the space-time asymmetry 
from the difference of the pseudoscalar meson 
decay constants determined from the temporal and spatial components 
of axial vector currents: 
\ben
R \equiv i \frac{\la 0|A_k^R|PS\ra}{\la 0|A_4^R|PS\ra}\cdot\frac{E}{|p_s|}, 
\een
which should become unity in the continuum limit.
Results are plotted in Figs.\ref{fig:splais}(c), (d).
We find that 
the asymmetry reaches about $10 \sim 20\%$ for the clover action
around $J/\Psi$ mass in the heavy-heavy system ,
while it is less than $7\%$ for the improved action.
These observations allow us to conclude that the improved action 
clearly reduces the errors caused by $m_Q a$.


\section{Conclusion and perspective}
\label{sec:conclusion}

We have proposed a 
relativistic $O(a)$ improvement to the heavy quarks on the lattice.
The idea is based on the relativistic on-shell improvement with
the finite $m_Q a$ corrections.
We have shown that the cutoff effects can be reduced to
$O((a\Lambda_{\rm QCD})^2)$ putting
the $(m_Q a)^n$ corrections
on the renormalization factors of the quark mass $Z_m$ and
wave function $Z_q$. 
Our relativistic approach has the strong point over the
nonrelativistic ones:  
the finer the lattice spacing becomes, the better the 
approach works. This is a desirable feature because
we can take the full advantage of configurations with
finer lattice spacing generated to control the cutoff
effects on the light hadron physics.

In the next step we have determined the $O(a)$ improvement coefficients, 
$\nu$, $r_s$, $\cb$ and $\ce$ in the quark action up to one-loop level
for the various improved gauge actions.
While $\nu$ and $r_s$ are determined from the quark propagator,
we use the on-shell quark-quark scattering amplitude for $\cb$ and $\ce$.
The $m_Q a$ dependences are examined by making the perturbative calculations
done in a $m_Q a$ dependent way.
Employing the conventional perturbative method with the fictitious gluon mass
as an infrared regulator we have shown that
the parameters $\nu$, $r_s$, $\cb$ and $\ce$ in the action are determined
free from the infrared divergences.
This is achieved if and only if the tree-level values for 
$\nu$, $r_s$, $\cb$ and $\ce$ are properly adjusted.

We have also made the $O(a)$ improvement of
the vector and the axial vector currents up to one-loop level.
Our calculation is carried out both for the heavy-heavy and the heavy-light cases
with the various gauge actions. It is explicitly shown that 
the renormalization and improvement coefficients of the heavy-light
vector current agree with those of the axial vector current,
once we impose the exact chiral symmetry for the light quark.

Given the $O(a)$ improved quark action and axial vector current,
we can check their effectiveness by investigating 
the restoration of the space-time symmetry for the
heavy-heavy and heavy-light meson systems.
We have focused on two quantities: the dispersion relation 
and the difference of the pseudoscalar meson
decay constants determined from the temporal and spatial components 
of axial vector currents. 
Our results show clear improvement for both quantities.

Up to now the $O(a)$ improvement of the massive Wilson quark action
works well: We do not encounter any theoretical contradiction 
in the improvement procedure of the quark action and the vector and axial
vector currents up to one-loop level, and
numerical studies clearly show the effectiveness of improvement.
These encouraging results lead us to the next step.
Our ongoing projects are the $O(a)$ improvement of the four-fermi
operators up to one-loop level and numerical simulations with
two- and three-flavors of dynamical quarks.
We also plan a detailed scaling study in quenched QCD
and a nonperturbative determination of the improvement coefficients.

\section*{Acknowledgements}
This work is supported in part by the Grants-in-Aid for
Scientific Research from the Ministry of Education, 
Culture, Sports, Science and Technology.
(Nos. 13135204, 14046202, 15204015, 15540251, 15740165.)


%

\end{document}